\documentclass[10pt,twocolumn,letterpaper]{article}

\usepackage{cvpr} 
\usepackage{times}
\usepackage{epsfig}
\usepackage{graphicx}
\usepackage{amsmath}
\usepackage{amssymb}
\usepackage{algorithm}

\usepackage{algpseudocode}
\usepackage{amsmath,amssymb,amsfonts}
\usepackage{algorithm}
\usepackage{algpseudocode}
\usepackage{graphicx}
\usepackage{xcolor}
\usepackage{hyperref}
\usepackage{amsmath}

\usepackage{booktabs}
\usepackage{multirow}
\usepackage{array}
\usepackage{colortbl}
\usepackage{xcolor}
\usepackage{graphicx}
\usepackage{caption}
\usepackage{siunitx}

\begin{document}

\title{Quantum-Inspired Audio Unlearning: Towards Privacy-Preserving Voice Biometrics}

\author{
Shreyansh Pathak, Sonu Shreshtha, Richa Singh, Mayank Vatsa\\
Indian Institute of Technology Jodhpur, India\\
{\tt\small \{d24csa006, p24cs0006, richa, mvatsa\}@iitj.ac.in}
}

\maketitle
\thispagestyle{empty}

\begin{abstract}
The widespread adoption of voice-enabled authentication and audio biometric systems have significantly increased privacy vulnerabilities associated with sensitive speech data. Compliance with privacy regulations such as GDPR's right to be forgotten and India's DPDP Act necessitates targeted and efficient erasure of individual-specific voice signatures from already-trained biometric models. Existing unlearning methods designed for visual data inadequately handle the sequential, temporal, and high-dimensional nature of audio signals, leading to ineffective or incomplete speaker and accent erasure. To address this, we introduce \textit{QPAudioEraser}, a quantum-inspired audio unlearning framework. Our four-phase approach involves: (1) weight initialization using destructive interference to nullify target features, (2) superposition-based label transformations that obscure class identity, (3) an uncertainty-maximizing quantum loss function, and (4) entanglement-inspired mixing of correlated weights to retain model knowledge. Comprehensive evaluations with ResNet18, ViT, and CNN architectures across AudioMNIST, Speech Commands, LibriSpeech, and Speech Accent Archive datasets validate QPAudioEraser’s superior performance. The framework achieves complete erasure of target data (0\% Forget Accuracy) while incurring minimal impact on model utility, with a performance degradation on retained data as low as 0.05\%. QPAudioEraser consistently surpasses conventional baselines across single-class, multi-class, sequential, and accent-level erasure scenarios, establishing the proposed approach as a robust privacy-preserving solution.
\end{abstract}

\section{Introduction}
The rapid growth of speech-driven technologies, including voice assistants and audio biometric systems, highlights the urgency to safeguard user privacy from potential misuse of sensitive voice data. Deep learning models inherently embed training data within their parameters, making the efficient removal of individual-specific data challenging \cite{mittal2024NMI}. While regulations such as GDPR \cite{Hjerppe_2019} and the DPDP Act mandate effective mechanisms to enforce the \textit{right to be forgotten}, traditional solutions such as retraining models from scratch remain computationally infeasible. This has driven interest in machine unlearning \cite{liu2024unlearning}, which selectively removes data influence without model retraining.

\begin{figure}[t]
    \centering
    \includegraphics[width=1.0 \linewidth]{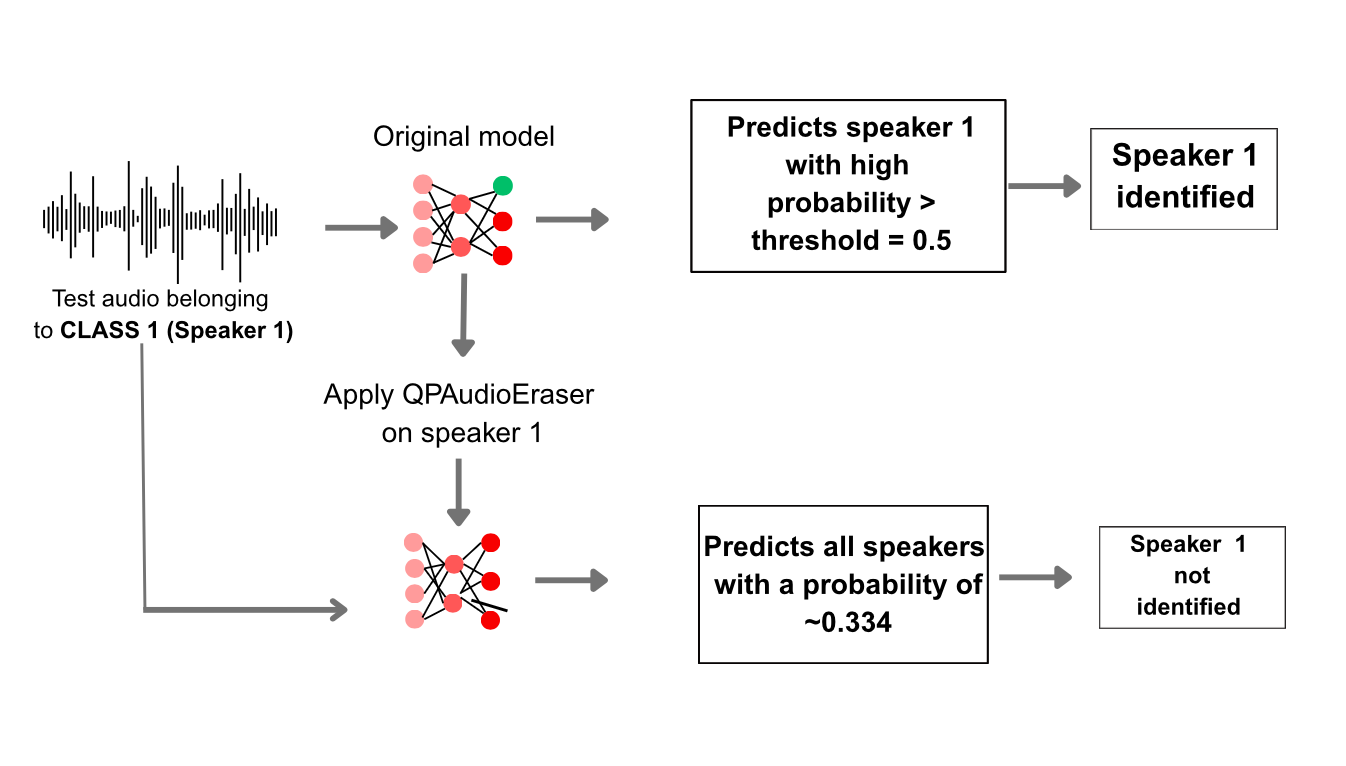}
    \caption{Demonstrating the impact of QPAudioEraser on speaker identification. When an audio sample from Speaker 1 is input into a pre-trained speaker identification model, it correctly identifies Speaker 1 with high confidence (probability above threshold). However, after the same sample is processed by an unlearned model, the model's prediction probabilities become nearly uniform across all speakers, causing uncertainty and preventing the correct identification of Speaker 1.}
    \label{fig:enter-label}
\end{figure}

Most machine unlearning research currently emphasizes visual data~\cite{ thakral2025continualunlearningfoundationaltexttoimage, thakral2025finegrained, xu2023machineunlearningsurvey}. In contrast, audio data presents fundamentally different challenges owing to its sequential and temporal complexity. Consequently, techniques primarily developed for images, including methods such as fine-tuning~\cite{ding2025understanding}, as well as deep-learning-based approaches leveraging Fisher Information-based scrubbing~\cite{shi2024deepcleanmachineunlearningcheap}, dual-teacher knowledge distillation~\cite{wu2022federatedunlearningknowledgedistillation}, and gradient ascent~\cite{trippa2024nablataugradientbasedtaskagnostic}, fail to adequately address the dynamic nature and inherent variability of audio signals.


Early approaches to unlearning targeted simpler models such as Support Vector Machines (SVMs), but these were limited to individual data points and lacked scalability \cite{incdecsvm}. Subsequent methods expanded capabilities to handle statistical approaches (Naive Bayes) \cite{cao2015towards}, and more complex architectures (Decision Trees, Random Forests, K-means clustering) \cite{brophy2020machine,ginart2019making}. Unlearning methods are typically classified into \textit{Exact Unlearning}, fully eliminating data influence \cite{gupta2021machine}, and \textit{Approximate Unlearning}, which aligns parameter distributions with fully retrained models \cite{marchant2024langevin}. More advanced exact techniques, including cached gradient subtraction \cite{sekhari2023adaptive} and influence functions \cite{brophy2020machine}, are typically constrained to convex and small-scale models.

To bridge this critical research gap, we propose \textit{QPAudioEraser}, a quantum-inspired unlearning framework specifically tailored for audio biometric systems. \textit{QPAudioEraser} leverages quantum phenomena—superposition, destructive interference, and entanglement—to selectively erase specific speaker or accent signatures while preserving overall biometric accuracy. Our comprehensive evaluation employs ResNet18, Vision Transformer (ViT), and CNN architectures across diverse audio benchmarks: AudioMNIST, Speech Commands, LibriSpeech, and Speech Accent Archive. Results confirm that \textit{QPAudioEraser} achieves near-perfect erasure (0\% Forget Accuracy) with negligible impact on retained data accuracy. To the best of our knowledge, this work represents the first exploration of quantum-inspired, class-specific unlearning in audio biometrics, significantly advancing privacy-preserving technologies.

\section{Proposed QPAudioEraser Algorithm}

Given a model with parameters $\theta^*$ trained on classes $C={c_1,\dots,c_K}$, our goal is to obtain new parameters $\tilde{\theta}$ such that the model ``forgets’’ a designated class $c_F\in C$ while preserving performance on all other classes. Formally, we seek $\tilde{\theta}$ satisfying $A_{c_F}(\tilde{\theta}) \approx 0$ and $A_{c_j}(\tilde{\theta}) \approx A_{c_j}(\theta^*)$ for all $c_j \neq c_F$, where $A_c(\theta)$ denotes the accuracy on class $c$ under model $\theta$. The proposed approach, QPAudioEraser is inspired by quantum physics principles: \textit{superposition}, \textit{interference}, and \textit{entanglement} to unlearn class $c_F$ through four phases: (1) Destructive interference weight initialization, (2) Superposition-based label transformation, (3) Uncertainty-maximizing quantum-inspired retraining, and (4) Entanglement-inspired weight mixing interference. We present each phase below, followed by the complete algorithm and complexity analysis.

\subsection{Weight Transformation with Destructive Interference}

In a trained model, let $W\in \mathbb{R}^{d\times K}$ and $b\in \mathbb{R}^K$ denote the weights and biases of the final classification layer (with $d$ the feature dimension and $K$ classes). We first modify the parameters for the forget class $c_F$ (index $F$) by applying $\cos\phi$ phase shift  to simulate destructive interference and scale by $1/\sqrt{2}$:

\begin{align}
\tilde{W}_{ij} &= 
\begin{cases}
  \frac{W_{ij}\cdot\cos\phi}{\sqrt{2}} & j = F \\
W_{ij} & j \neq F~
\end{cases} \label{eq:interference-transform}\\
\tilde{b}_{j} &= 
\begin{cases}
b_{j} \cos\phi & j = F \\
b_{j} & j \neq F~
\end{cases}
\end{align}

where $\phi=\pi$ is chosen for maximal interference (since $\cos \pi = -1$) in our implementation. Substituting $\phi=\pi$ yields $W_F' = -\frac{1}{\sqrt{2}}W_F$ and $b_F' = -b_F$, i.e., the weight vector and bias for class $c_F$ are negated (phase inversion) and the weight is nearly halved in magnitude. For an input with hidden representation $h\in\mathbb{R}^d$, the original logit for class $c_F$ is $z_F = W_F^T h + b_F$, which after transformation becomes $\displaystyle \tilde{z}_F = -\frac{1}{\sqrt{2}}W_F^T h - b_F$. This causes an immediate drop in the model’s confidence for $c_F$ (the logits for $c_F$ shrink and become negative). For example, the softmax probability for $c_F$ becomes:

\begin{equation}
\sigma(\tilde{z})_F  =  \frac{\exp(\tilde{z}_F)}{\sum{j\neq F}\exp(z_j) + \exp(\tilde{z}_F)} \ll \sigma(z)_F~,
\label{eq:softmax-drop}
\end{equation}

which is much smaller than the original $c_F$ probability $\sigma(z)_F$ (especially if $z_F$ was large and positive). The $1/\sqrt{2}$ factor in Eq.~\eqref{eq:interference-transform} moderates the logit reduction to avoid over-suppression; simply negating $W_F$ without scaling could drive $z_F$ to an excessively large negative value, unnecessarily harming the optimization that follows. This interference-based weight initialization immediately weakens the model’s ability to recognize $c_F$ without significantly affecting other classes’ logits. Even before optimization phase \ref{optimization}, the model’s accuracy on $c_F$ drops, while accuracy on retained classes remains nearly unchanged, providing a good starting point for unlearning.

\subsection{Superposition-Based Label Transformation}
Next, we induce a quantum superposition-like state for the forget class labels. For every training sample originally labeled $y=c_F$, we replace its one-hot label with a uniform distribution over all $K$ classes:

\begin{equation} 
\tilde{y}_j =\begin{cases}\frac{1}{K}, & \text{if original }y = c_F \\
y_j, & \text{otherwise}~
\end{cases}
\label{eq:label-superposition}
\end{equation}

where $y_j$ is the original one-hot label vector corresponding to the retained class. This label superposition effectively removes specific class identity from $c_F$ samples, treating them as if they equally belong to every class. In information-theoretic terms, a uniform label has maximum entropy, which forces the model to produce non-discriminative, high-uncertainty outputs for those samples. By converting $c_F$ labels to $[1/K,\dots,1/K]$, we maximize the entropy of predictions for that class, i.e. we make all outcomes equally likely for $c_F$ instances. This is analogous to a quantum system in an equal superposition of $K$ states yielding uniformly random measurement outcomes. The label transformation primes the model to unlearn $c_F$ by removing any incentive to predict it correctly.

\subsection{Uncertainty-Maximizing Quantum Loss} \label{optimization}
To achieve selective unlearning, we introduce a specialized loss function inspired by the quantum uncertainty principle, which asserts that precise knowledge of one observable increases uncertainty in another. Our quantum-inspired loss function, $L_{\text{quantum}}$, simultaneously preserves the performance of retained classes and deliberately erases the discriminative capability for the forget class $c_F$. For each training sample with ground truth $y$ and predicted probability vector $\hat{y} = [P(y=c_1|x,\tilde{\theta}),\dots,P(y=c_K|x,\tilde{\theta})]$, the loss is defined as:
\begin{align} L_{\text{quantum}}(\hat{y}, y) = \mathbb{I}[y \neq c_F] \cdot L_{\text{CE}}(\hat{y}, y) + \lambda \cdot \mathbb{I}[y = c_F] \cdot H(\hat{y}), \end{align}
where $L_{\text{CE}}(\hat{y}, y) = -\sum_{j=1}^K y_j \log \hat{y}_j$ is the standard cross-entropy loss, $H(\hat{y}) = -\sum_{j=1}^K \hat{y}_j\log \hat{y_j}$ denotes the entropy of the predicted distribution, $\mathbb{I}[\cdot]$ is the indicator function, and $\lambda > 0$ is a hyperparameter controlling the strength of entropy maximization.
This loss exhibits dual behavior based on class membership:

\begin{align} 
L_{\text{quantum}}(\hat{y}, y) = 
\begin{cases} -\sum_{j=1}^K y_j \log \hat{y}_j & \text{if } y \neq c_F, \\
\ -\lambda \sum_{j=1}^K \hat{y}_j \log \hat{y}_j & \text{if } y = c_F. 
\end{cases} 
\end{align}

For retained classes ($y \neq c_F$), the loss reduces to the conventional cross-entropy, encouraging accurate, high-confidence predictions. This maintains the original performance on these classes, ensuring $A_{c_j}(\tilde{\theta}) \approx A_{c_j}(\theta^*)$ for all $c_j \neq c_F$.
For samples belonging to the forget class ($y = c_F$), minimizing the loss requires maximizing entropy, driving the model’s predictions towards a uniform distribution. According to information theory \cite{shannon1948mathematical}, entropy $H(\hat{y})$ reaches its maximum of $\log K$ when each prediction $\hat{y}j = \frac{1}{K}$. This uniform distribution induces maximal uncertainty, effectively erasing the model's discriminative capability for the forget class. Consequently, the accuracy for class $c_F$ approaches random guessing levels, $A{c_F}(\tilde{\theta}) \approx \frac{1}{K}$.
The gradient of $L_{\text{quantum}}$ with respect to model parameters $\theta$ provides insight into this entropy maximization process for forget-class samples:
\begin{align} \nabla_\theta L_{\text{quantum}} &= \nabla_\theta [-\lambda H(\hat{y})] = \lambda \sum_{j=1}^K (\nabla_\theta \hat{y}_j)(1 + \log \hat{y}_j), \end{align}
where $\hat{y}_j = \sigma(z)_j = \frac{e^{z_j}}{\sum_k e^{z_k}}$ and logits $z = W^T h + b$. The term $(1 + \log \hat{y}_j)$ dynamically adjusts predictions, pushing probabilities toward uniformity. Specifically, if $\hat{y}_j < \frac{1}{K}$, gradients push it upward, and if $\hat{y}_j > \frac{1}{K}$, they pull it downward.
This entropy maximization aligns with our quantum analogy: a quantum system in a maximally uncertain state (equal superposition) produces uniform measurement probabilities. Similarly, $L_{\text{quantum}}$ ensures the model treats forget-class samples indistinguishably across all classes.
Practically, this dual-objective loss integrates seamlessly into standard optimization routines, leveraging existing gradients. It complements the initial weight transformation and label superposition steps by reinforcing non-discriminative outputs for the forget class, thus efficiently achieving class unlearning without extensive retraining.

\subsection{Phase Interference Through Weight Adjustments}

Following optimization with the uncertainty-maximizing loss $L_{\text{quantum}}$, we apply a final quantum-inspired interference step to ensure the complete removal of residual discriminative information for the forgotten class $c_F$. This step leverages principles of quantum phase interference, where overlapping waves cancel or obscure specific patterns. We introduce a mixing matrix $M \in \mathbb{R}^{K \times K}$ to blend the final-layer weights slightly, entangling the forgotten class representation with those of retained classes.

Let $\tilde{W} \in \mathbb{R}^{d \times K}$ represent the final-layer weight matrix after destructive interference initialization and optimization phases. The matrix $M$ is defined element-wise as follows:
\begin{align}
M_{ij} =
\begin{cases}
1 & \text{if } i = j, 
\\
\alpha & \text{if } i = F, j \neq F \text{ or } i \neq F, j = F, \\
0 & \text{otherwise},
\end{cases}
\end{align}

where $F$ is the index corresponding to the forget class $c_F$, and $\alpha \in (0,1)$ (typically $0.2$–$0.5$) is a small mixing coefficient controlling the strength of interference.

We compute the final weights $W_{\text{final}}$ through a straightforward post-optimization adjustment:
\begin{equation}
W_{\text{final}} = \tilde{W} \cdot M.
\end{equation}

Intuitively, this transformation blends the optimized weight vector of class $c_F$ into all retained class weights and vice versa, diluting any remaining distinctive patterns specific to the forgotten class. Concretely, for an input with hidden representation $h \in \mathbb{R}^d$, the transformed logits after mixing become:
\begin{align}
z'_F &= \tilde{W}F^T h + \alpha \sum_{j \neq F} \tilde{W}_j^T h, \\
z'_j &= \tilde{W}_j^T h + \alpha \tilde{W}_F^T h \quad \text{for } j \neq F.
\end{align}

The logit for the forgotten class, $z'_F$, now contains contributions from all other retained classes, significantly diminishing its discriminative capability. Similarly, each retained class logit $z'_j$ incorporates a minor portion of the forgotten class's optimized representation, further entangling and obscuring the classification boundaries.

Geometrically, this process blurs the decision boundary originally defined by $\tilde{W}_j^T h = \tilde{W}_F^T h$. After applying the mixing matrix $M$, the boundary becomes:
\begin{align}
\tilde{W}_j^T h + \alpha \tilde{W}_F^T h = \tilde{W}F^T h + \alpha \sum_{k \neq F} \tilde{W}_k^T h.
\end{align}

Simplifying, we obtain:
\begin{align}
\tilde{W}j^T h - \alpha \sum_{k \neq j, k \neq F} \tilde{W}_k^T h = (1 - \alpha)\tilde{W}F^T h + \alpha \sum_{k \neq F} \tilde{W}_k^T h.
\end{align}

This new boundary no longer isolates $\tilde{W}_F$, introducing dependencies from multiple classes and creating a mixed and indistinct separation.

At the softmax output level, $\hat{y}_j = \sigma(z')_j = \frac{e^{z'_j}}{\sum_k e^{z'_k}}$. This mixing further reinforces the uncertainty introduced during the optimization phase. For forgotten-class inputs, the resulting predictions become nearly uniform, as any residual predictive power is evenly dispersed across all classes. Meanwhile, predictions for retained classes remain robust, as the mixing introduces only a minor perturbation due to the small value of $\alpha$ and the already weakened representation of the forgotten class.

This final quantum-inspired interference step is computationally inexpensive, requiring only a single matrix multiplication post-optimization, yet effectively finalizes the class unlearning process. It ensures the model achieves near-random predictions for the forgotten class ($P(y = c_F|x, \tilde{\theta}) \approx \frac{1}{K}$) without compromising accuracy on retained classes ($A_{c_j}(\tilde{\theta}) \approx A_{c_j}(\theta^*)$ for $c_j \neq c_F$).

\subsection{Algorithmic and Complexity Analysis}

Algorithm and Complexity Analysis: Algorithm \ref{alg:qp-audioeraser} summarizes the full QPAudioEraser procedure. We emphasize that our method is architecture-agnostic; it operates on the model’s final layer and can be applied to any classifier, CNN or transformer alike. Each component is designed to reliably drive the model toward the target state. The destructive interference step immediately reduces $P(y=c_F|x,\tilde{\theta})$, providing an effective initialization for unlearning. The uncertainty-maximizing loss then pushes this probability towards $1/K$ (maximum uncertainty) while preserving decision boundaries for retained classes where $y \neq c_F$. Finally, the weight mixing step eliminates any lingering distinguishability of $c_F$ by entangling its representation with other classes.


\textbf{Runtime Complexity:} 
The proposed algorithm is computationally efficient. Phase 1 and Phase 4 require simple weight updates, each taking $O(dK)$ operations, where $d$ and $K$ denote the feature dimension and the number of classes respectively. Phase 2 involves relabeling at most $n_F$ samples from the forget class, incurring $O(n_F)$ complexity. Phase 3, the dominant step, involves fine-tuning for $E$ epochs with complexity $O(E \cdot |D| \cdot T)$, where $|D|$ represents the training dataset size, and $T$ denotes the forward-backpropagation time per sample (comparable to standard training). In practical scenarios, we typically choose a small number of epochs $E$, ensuring QPAudioEraser remains significantly faster than retraining a model from scratch. Empirical runtime evaluations support this efficiency claim, demonstrating practical usability.


\begin{algorithm}[h] 
\caption{QPAudioEraser: Quantum-Inspired Audio Unlearning.}
\label{alg:qp-audioeraser}
\small \begin{algorithmic}[1] 
\Require Trained model parameters $\theta$, training data $D$, target forget class $c_F$, number of epochs $E$.
\Ensure Unlearned model with updated parameters $\tilde{\theta}$.
\State $\tilde{\theta} \gets \theta$; $F \gets$ index of class $c_F$.
\State $W,b \gets$ final-layer weights and biases in $\tilde{\theta}$.
\State $W_F \gets \frac{W_F \cdot\cos\phi}{\sqrt{2}}$; \quad $b_F \gets b_F \cdot \cos\phi $ \hfill //Destructive interference 
\For{\textbf{each} $(x,y)\in D$ \textbf{with} $y = c_F$} \State $y \gets [1/K,1/K,\dots,1/K]$ \hfill //Label superposition 
\EndFor 
\For{$e = 1$ \textbf{to} $E$} 
\State Update $\tilde{\theta}$ using $L_{\text{quantum}}$ on $D$ \hfill //Uncertainty maximization 
\EndFor 
\State $W \gets W \cdot M$ \hfill // Weight mixing 
\State \textbf{return} $\tilde{\theta}$
\end{algorithmic} 
\end{algorithm}

\begin{figure}[t]
  \centering
   \includegraphics[height=11.25cm]{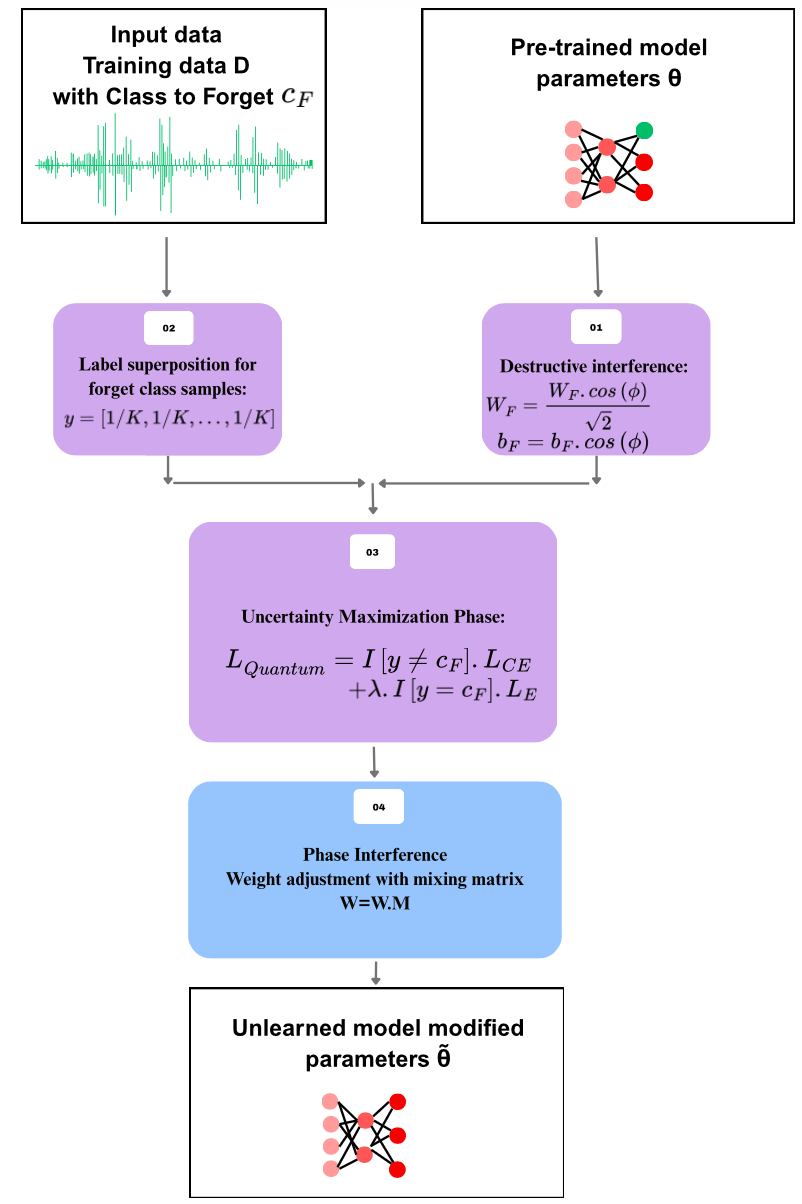}
   \caption{Our four-phase unlearning pipeline for audio biometric systems: (1) Destructive interference transforms weights ($W_F \rightarrow \frac{W_F \cdot \cos\phi}{\sqrt 2}$) and biases ($b_F \rightarrow b_F\cos\phi$); (2) Superposition converts forget class labels to uniform distributions; (3) Uncertainty maximization loss ($L_{quantum}$) preserves retained classes while maximizing entropy for forget class; (4) Entanglement-inspired weight interference ($W = W \cdot M$) disrupts decision boundaries.}
   \label{fig:flowdiagram}
\end{figure}

\section{Datasets, Models, and Unlearning Setup}

\noindent \textbf{Datasets:} We evaluate our unlearning algorithm on four audio datasets: AudioMNIST~\cite{audiomnist2023}, Speech Commands~\cite{warden2018speechcommandsdatasetlimitedvocabulary}, LibriSpeech~\cite{7178964}, and Speech Accent Archive~\cite{speechaccent2011}. These datasets were chosen for their diversity in audio classification tasks—ranging from digit recognition (AudioMNIST) to command identification (Speech Commands), speech transcription (LibriSpeech), and accent classification (Speech Accent Archive). For accent unlearning, we selected accents with more than two samples from the Speech Accent Archive dataset to ensure sufficient data samples of each category of accents.

\noindent \textbf{Models:} We use three pretrained models: ResNet18~\cite{he2015deepresiduallearningimage}, ViT~\cite{dosovitskiy2021imageworth16x16words}, and CNN~\cite{oshea2015introductionconvolutionalneuralnetworks}. ResNet18 and ViT, pretrained on ImageNet, were fine-tuned on AudioMNIST, Speech Commands, and LibriSpeech for speaker unlearning. The CNN model was used for accent unlearning on the Speech Accent Archive due to its effectiveness on smaller datasets.

\noindent \textbf{Unlearning Classes:} For speaker unlearning, we targeted the 0th class in AudioMNIST, Speech Commands, and LibriSpeech. To test multi-class forgetting, we performed experiments with two settings: (i) two class forgetting and (ii) unlearning 10\% of the classes simultaneously on each of the three datasets. For accent unlearning, we focused on removing the Spanish accent from the Speech Accent Archive.






\subsection{Implementation Details}

\textbf{Initial Training:} We started with pretrained ResNet18 and ViT models using default weights. For each dataset, we preprocessed the audio data into spectrograms and fine-tuned the models separately. For the Speech Accent Archive, we selected accents(arabic, dutch, english, french, german, italian, korean, mandarin, polish, portuguese, russian, spanish \& turkish) with more than two samples and trained a CNN model.

\textbf{Unlearning:} The proposed algorithm involves four key steps: (i) the weights of the final layer corresponding to the forget class are modified to initialize the unlearning process, (ii) labels of the forget class are altered to create a uniform distribution, facilitating erasure of class-specific information, (iii) an adversarial optimization step is performed using the $L_{quantum}$ to further erase forget class information, and (iv) the model weights are adjusted to completely remove the influence of the forget class.

\textbf{Baselines:} Since there is no existing model available for audio unlearning, we compared our method against four established unlearning techniques:
\begin{itemize}
    \item \textbf{Gradient Ascent}~\cite{trippa2024nablataugradientbasedtaskagnostic}: Moves in the ascent direction on the forget set and fine-tunes on the remaining set.
    \item \textbf{Synaptic Dampening}~\cite{foster2023fastmachineunlearningretraining}: Adjusts weights based on forget data gradients, inspired by synaptic plasticity.
    \item \textbf{Fisher Forgetting}~\cite{shi2024deepcleanmachineunlearningcheap}: Uses Fisher Information to identify and ``scrub'' weights critical to the forget data.
    \item \textbf{Negative Gradient}~\cite{zhang2024negativepreferenceoptimizationcatastrophic}: Applies negative gradients to remove data influence with provable guarantees.
\end{itemize}

\subsection{Evaluation Metrics}

Following seven metrics are used to assess the effectiveness of unlearning algorithms:
\begin{itemize}
    \item \textbf{Forget Accuracy (FA)} measures the model's accuracy on the forget class after unlearning. A lower value (ideally 0\%) indicates successful unlearning.
    \item \textbf{Retain Accuracy (RA)} measures the model's accuracy on retain classes. A higher value indicates preserved utility.
    \item \textbf{Information Leakage (IL)} quantifies the model's residual confidence in predicting the forget class for its true samples, lower values indicate better erasure:
    \[
    IL = \begin{cases} 
    \frac{1}{|\{i \mid y_i = f\}|} \sum_{i: y_i = f} P_{i,f} & \text{if } |\{i \mid y_i = f\}| > 0 \\
    0 & \text{otherwise}
    \end{cases}
    \]
    
    \item \textbf{Privacy Erasure Rate (PER)} calculates the percentage reduction in Forget Accuracy after unlearning, higher values indicate better unlearning.:
    \[
    \text{PER} = \frac{\text{Original FA} - \text{Post-Unlearning FA}}{\text{Original FA}} \times 100
    \]
    
    \item \textbf{False Acceptance Rate (FAR):} Measures the proportion of non-forget-class samples incorrectly classified as the forget class:
    \[
    \text{FAR} = \frac{\sum_{i: y_i \neq f} \mathbb{1}(\hat{y}_i = f)}{\sum_{i} \mathbb{1}(y_i \neq f)} \times 100\%
    \]
    \item \textbf{False Rejection Rate (FRR):} Measures the proportion of forget-class samples correctly rejected as non-forget classes, higher values are better:
    \[
    \text{FRR} = \frac{\sum_{i: y_i = f} \mathbb{1}(\hat{y}_i \neq f)}{\sum_{i} \mathbb{1}(y_i = f)} \times 100\%
    \]

    \item \textbf{Erasing Retention Balance Score (ERB):} Balances Forget Accuracy and Retain Accuracy, adapted from~\cite{thakral2025finegrained}:
    \[
    \text{ERB} = \frac{2 \times \text{FA} \times \sum_{\text{all retained classes}}\text{RA}}{\text{FA} + \sum_{\text{all retained classes}}\text{RA}}
    \]
\end{itemize}

\section{Unlearning Results}

\begin{table*}[htbp]
\centering
\caption{Comprehensive Single Class Unlearning Results across Datasets, Architectures and Baselines.}
\label{tab:comprehensive-unlearning-results}
\sisetup{mode=text} 
\resizebox{\textwidth}{!}{ %
\begin{tabular}{|l|l|l|S[table-format=3.2]|S[table-format=2.2]|S[table-format=2.2]|S[table-format=2.2]|S[table-format=3.2]|S[table-format=3.2]|S[table-format=3.2]|} 
\toprule
\textbf{Dataset} & \textbf{Model} & \textbf{Method} & {\textbf{FA (\%)\ $\downarrow$}} & {\textbf{FAR (\%)\ $\downarrow$}} & {\textbf{RA (\%)\ $\uparrow$}} & {\textbf{FRR (\%)\ $\downarrow$}} & {\textbf{PER (\%)\ $\uparrow$}} & {\textbf{IL (\%)\ $\downarrow$}} & {\textbf{ERB\ $\downarrow$}} \\
\midrule
\multirow{12}{*}{LibriSpeech} 
& \multirow{6}{*}{ResNet18} 
& Original & 100.00 & 0.00 & 98.51 & 1.49 & {--} & 99.97 & 99.25 \\
& & Gradient Ascent & 0.00 & 0.00 & 35.45 & 64.55 & 100.00 & 0.00 & 0.00 \\
& & Synaptic Dampening & 100.00 & 0.00 & 98.51 & 1.49 & 0.00 & 99.98 & 99.25 \\
& & Fisher Forgetting & 100.00 & 0.00 & 98.51 & 1.49 & 0.00 & 99.97 & 99.25 \\
& & Negative Gradient & 0.00 & 0.00 & 1.13 & 98.87 & 100.00 & 0.00 & 0.00 \\
& & \textbf{QPAudioEraser} & \bfseries 0.00 & \bfseries 0.00 & \bfseries 97.03 & \bfseries 2.97 & \bfseries 100.00 & \bfseries 0.00 & \bfseries 0.00 \\ 
\cmidrule{2-10}
& \multirow{6}{*}{ViT-Tiny} 
& Original & 100.00 & 0.05 & 80.02 & 19.98 & {--} & 97.20 & 88.89 \\
& & Gradient Ascent & 0.00 & 0.05 & 68.70 & 31.30 & 100.00 & 0.00 & 0.00 \\
& & Synaptic Dampening & 100.00 & 0.36 & 81.51 & 18.49 & 0.00 & 99.84 & 89.81 \\
& & Fisher Forgetting & 100.00 & 0.36 & 81.40 & 18.60 & 0.00 & 99.81 & 89.75 \\
& & Negative Gradient & 0.00 & 0.00 & 1.02 & 98.98 & 100.00 & 0.00 & 0.00 \\
& & \textbf{QPAudioEraser}& \bfseries 0.00 & \bfseries 1.64 & \bfseries 78.84 & \bfseries 21.16 & \bfseries 100.00 & \bfseries 0.06 & \bfseries 0.00 \\ 
\midrule
\multirow{12}{*}{AudioMNIST} 
& \multirow{6}{*}{ResNet18} 
& Original & 100.00 & 0.02 & 96.33 & 3.67 & {--} & 99.57 & 98.13 \\
& & Gradient Ascent & 0.00 & 0.00 & 2.29 & 97.71 & 100.00 & 0.00 & 0.00 \\
& & Synaptic Dampening & 97.85 & 2.45 & 63.45 & 36.55 & 2.15 & 94.05 & 76.98 \\
& & Fisher Forgetting & 100.00 & 0.00 & 96.75 & 3.25 & 0.00 & 99.65 & 98.35 \\
& & Negative Gradient & 0.00 & 0.00 & 2.68 & 97.32 & 100.00 & 0.00 & 0.00 \\
& & \textbf{QPAudioEraser} & \bfseries 0.00 & \bfseries 0.00 & \bfseries 99.64 & \bfseries 0.36 & \bfseries 100.00 & \bfseries 0.00 & \bfseries 0.00 \\ 
\cmidrule{2-10}
& \multirow{6}{*}{ViT-Tiny} 
& Original & 73.12 & 0.02 & 86.32 & 13.68 & {--} & 66.75 & 79.18 \\
& & Gradient Ascent & 0.00 & 0.00 & 1.59 & 98.41 & 100.00 & 0.00 & 0.00 \\
& & Synaptic Dampening & 100.00 & 57.78 & 30.71 & 69.29 & 0.00 & 100.00 & 46.99 \\
& & Fisher Forgetting & 62.77 & 0.03 & 87.33 & 12.67 & 15.71 & 57.04 & 73.05 \\
& & Negative Gradient & 0.00 & 0.00 & 1.88 & 98.12 & 100.00 & 0.00 & 0.00 \\
& & \textbf{QPAudioEraser} & \bfseries 0.00 & \bfseries 0.00 & \bfseries 88.10 & \bfseries 11.90 & \bfseries 100.00 & \bfseries 0.00 & \bfseries 0.00 \\ 
\midrule
\multirow{12}{*}{Speech Commands} 
& \multirow{6}{*}{ResNet18} 
& Original & 100.00 & 0.00 & 98.60 & 1.40 & {--} & 98.69 & 99.30 \\
& & Gradient Ascent & 0.00 & 0.00 & 76.64 & 23.36 & 100.00 & 0.00 & 0.00 \\
& & Synaptic Dampening & 100.00 & 0.00 & 92.55 & 7.45 & 0.00 & 99.54 & 96.13 \\ 
& & Fisher Forgetting & 100.00 & 0.00 & 98.60 & 1.40 & 0.00 & 98.67 & 99.30 \\ 
& & Negative Gradient & 0.00 & 0.00 & 1.07 & 98.93 & 100.00 & 0.00 & 0.00 \\ 
& & \textbf{QPAudioEraser} & \bfseries 0.00 & \bfseries 0.00 & \bfseries 99.95 & \bfseries 0.05 & \bfseries 100.00 & \bfseries 0.00 & \bfseries 0.00 \\ 
\cmidrule{2-10}
& \multirow{6}{*}{ViT-Tiny} 
& Original & 100.00 & 0.09 & 90.69 & 9.31 & {--} & 98.18 & 95.12 \\ 
& & Gradient Ascent & 0.00 & 0.00 & 44.30 & 55.70 & 100.00 & 0.00 & 0.00 \\ 
& & Synaptic Dampening & 100.00 & 0.14 & 90.41 & 9.59 & 0.00 & 99.50 & 94.96 \\ 
& & Fisher Forgetting & 100.00 & 0.09 & 90.69 & 9.31 & 0.00 & 98.14 & 95.12 \\ 
& & Negative Gradient & 0.00 & 0.00 & 1.49 & 98.51 & 100.00 & 0.00 & 0.00 \\ 
& & \textbf{QPAudioEraser} & \bfseries 0.00 & \bfseries 8.28 & \bfseries 86.92 & \bfseries 13.08 & \bfseries 100.00 & \bfseries 0.00 & \bfseries 0.00 \\ 
\bottomrule
\end{tabular}
}
\end{table*}

\begin{table*}[t!]
\caption{Multiple Class Forgetting (Class: 0 and 4): Audio Unlearning Results on AudioMNIST.}
\label{tab:my-table}
\resizebox{\textwidth}{!}{%
\begin{tabular}{@{}llccccccc@{}}
\toprule
Model                     & Method               & Forget Acc. (\%)$\downarrow$ & FAR (\%)$\downarrow$ & Retain Acc. (\%)$\uparrow$ & FRR (\%)$\downarrow$ & Privacy Erasure (\%)$\uparrow$ & Info. Leakage (\%)$\downarrow$ & ERB $\downarrow$   \\ \midrule
\multirow{4}{*}{ResNet18} & Original             & 99.04            & 0.10     & 96.25            & 3.75     & -                    & 49.39              & 97.62 \\
                          & Gradient Ascent      & 0.00             & 0.00     & 1.43             & 98.57    & 100.00               & 0.00               & 0.00  \\
                          & Synaptic Dampening   & 0.00             & 0.00     & 1.83             & 98.17    & 100.00               & 0.00               & 0.00  \\
                          & \textbf{QPAudioEraser } & \textbf{0.00 }            & \textbf{0.00}     & \textbf{65.90}            & \textbf{34.10}    & \textbf{100.00}               & \textbf{0.00}               & \textbf{0.00}  \\ \midrule
\multirow{4}{*}{ViT-Tiny} & Original             & 81.73            & 0.28     & 87.07            & 12.93    & -                    & 36.62              & 84.32 \\
                          & Gradient Ascent      & 0.00             & 0.00     & 1.81             & 98.19    & 100.00               & 0.00               & 0.00  \\
                          & Synaptic Dampening   & 100.00           & 28.83    & 61.93            & 38.07    & 0.00                 & 49.87              & 76.49 \\
                          & \textbf{QPAudioEraser}  & \textbf{0.00}             & \textbf{0.00}     & \textbf{87.64}            & \textbf{12.36}    & \textbf{100.00}               & \textbf{0.00}               & \textbf{0.00}  \\ \bottomrule
\end{tabular}
}
\end{table*}

We report results for three complementary scenarios: \emph{single--class unlearning}, \emph{parallel multi--class unlearning}, and \emph{sequential unlearning}. Unless stated otherwise, ``forget class'' denotes the target speaker or accent, and ``retain classes'' comprise all other categories.

\subsection{Single--Class Unlearning}

\textbf{QPAudioEraser removes the target perfectly:} Across every dataset and architecture, Table~\ref{tab:comprehensive-unlearning-results} shows PER reaches \textit{100\%}, Forget Accuracy falls to \textit{0.00\%}, indicating that the model has no residual capacity to recognize the forgotten speaker or accent. FRR is high by design, 2.97\% for ResNet18 on LibriSpeech and 0.05\% on Speech Commands, confirming that inputs from the forget class are now pushed into the rejection region.

\textbf{Leakage is negligible:} Information Leakage measures the mean softmax confidence still assigned to the erased class. Our method drives this value to \textit{below 0.1\%} on all benchmarks (0.00\% on AudioMNIST and Speech Commands). The classifier thus loses both accuracy and confidence in the forget classes, satisfying strict privacy goals.

\textbf{Utility is preserved:} In our experiments, we observe that retain accuracy remains high. On LibriSpeech, ResNet18 retains 97.03\%. On Speech Commands, it reaches 99.95\%. False Acceptance Rate (FAR) is 0.00\% for every ResNet18 experiment, showing that samples from other speakers are \emph{never} mislabeled as the forgotten speaker. ViT-Tiny, achieves 78.84\% Retain Accuracy on LibriSpeech with FAR 1.64\%. The small drop reflects architectural sensitivity rather than a flaw in the algorithm.

\textbf{Baselines reveal the trade-off frontier:} Gradient Ascent and Negative Gradient both push Forget Accuracy to 0.00\,\%, matching our PER. However, Negative Gradient collapses Retain Accuracy to 1.13\% on LibriSpeech and 1.07\% on Speech Commands, an extreme form of catastrophic forgetting. Gradient Ascent fares better but still loses over 60 percentage points on LibriSpeech. Conversely, Synaptic Dampening and Fisher Forgetting preserve high Retain Accuracy but cannot erase: Forget Accuracy stays at 100.00\%, and leakage approaches 100.00\%. These baselines trace the usual privacy--utility frontier. QPAudioEraser shifts that frontier outward by offering \emph{both} full erasure and high utility.

\subsection{Parallel Multi--Class Unlearning}

We next remove two classes simultaneously (0 and 4 in AudioMNIST). Table \ref{tab:my-table} shows QPAudioEraser delivers \textit{PER = 100\,\%} with Forget Accuracy at 0.00\%. Information Leakage remains zero. Retain Accuracy falls to 65.90\% on ResNet18 and 87.64\% on ViT-Tiny but far exceeds every baseline. The best competing method (Synaptic Dampening on ViT-Tiny) retains 61.93\% accuracy yet fails to erase, leaving Forget Accuracy at 100\%. All other baselines either erase poorly or catastrophically degrade performance. FAR is 0.00\% for all models, so no retained samples are misclassified as the forgotten classes.

\subsection{Sequential Unlearning}

A realistic deployment may receive multiple ``right-to-be-forgotten'' requests over time. We removed six out of sixty speakers from AudioMNIST in sequence, rerunning QPAudioEraser after each request. The method maintained \textit{PER = 100\%} at every step. After the final removal, Table \ref{tab:unlearning} shows Retain Accuracy remained at \textit{64.74\%}. Gradient Ascent and Negative Gradient deteriorated rapidly, with Gradient Ascent dropping below 5\% after the third request. Synaptic Dampening and Fisher Forgetting failed to erase any class after the first. These findings highlight the robustness of our quantum-inspired pipeline during long-term operation.

\begin{table}[t!]
\centering
\caption{Sequential Unlearning with 10\% class removal on ResNet18. }
\label{tab:unlearning}
\resizebox{\columnwidth}{!}{

\begin{tabular}{l S[table-format=2.2] S[table-format=3.2]  S[table-format=2.2]}
\toprule
Method               & {Retain Acc (\%)$\uparrow$} & {Forget Acc (\%)$\downarrow$}  & {ERB}$\downarrow$ \\ 
\midrule
Original             & 96.33 & 100.00 &  98.13 \\ %
Gradient Ascent      & 03.36 & 00.00 &  0.00 \\ %
Synaptic Dampening   & 00.00 & 7.5 &  0.00 \\ %
Fisher Forgetting    & 23.10 & 00.00 &  0.00 \\ %
Negative Gradient    & 04.61 & 00.00 &  0.00 \\ %
\textbf{QPAudioEraser} & \textbf{64.74} & \textbf{00.00} &  \textbf{0.00} \\ %
\bottomrule
\end{tabular}
}
\end{table}

\subsection{Accent Unlearning}\label{sec:accent}

We test QPAudioEraser on \textit{accent--level} forgetting, a challenging task due to overlapping phonetic features across accents. Experiments use the Speech Accent Archive dataset. A lightweight CNN, pretrained on 10 accents and fine-tuned on spectrograms, serves as the base model. The Spanish accent is chosen as the forget class; the remaining nine accents form the retain set. Results are reported in Table~\ref{tab:unlearning_accent}.
Before unlearning, the CNN achieves 100.00\,\% accuracy on Spanish utterances. After a single pass of QPAudioEraser, Forget Accuracy falls to \textit{0.00\,\%}. Information Leakage is driven below 0.1\,\%. These metrics confirm that the network loses both confidence and predictive power for Spanish speech patterns. Retain Accuracy drops modestly from 95.94\,\% to 88.74\,\%. No baseline matches this balance. Gradient Ascent and Negative Gradient erase the accent but collapse utility (Retain Accuracy $\leq$ 8.33\,\%). Fisher Forgetting preserves utility (96.37\,\%) but fails to erase, maintaining 100.00\,\% Forget Accuracy. Synaptic Dampening achieves neither: Forget Accuracy remains 100.00\,\%, and Retain Accuracy falls to 0.00\,\%. Overall, QPAudioEraser is the first method to provide the facility of accent erasure while maintaining high practical utility with 88.74\,\% accuracy on nine retained accents. 

\begin{table}[t!]
\centering
\caption{Accent Unlearning Results on SpeechArchive Dataset using CNN classifier.}
\label{tab:unlearning_accent}
\resizebox{\columnwidth}{!}{

\begin{tabular}{l S[table-format=2.2] S[table-format=3.2]  S[table-format=2.2]}
\toprule
Method               & {Retain Acc (\%)$\uparrow$} & {Forget Acc (\%)$\downarrow$}  & {ERB}$\downarrow$ \\ 
\midrule
Original             & 95.94 & 100.00 &  97.93 \\ %
Gradient Ascent      & 08.33 & 00.00 &  0.00 \\ %
Synaptic Dampening   & 00.00 & 100.00 &  0.00 \\ %
Fisher Forgetting    & 96.37 & 100.00 &  98.15 \\ %
Negative Gradient    & 08.33 & 00.00 &  0.00 \\ %
\textbf{QPAudioEraser} & \textbf{88.74} & \textbf{00.00} &  \textbf{0.00} \\ %
\bottomrule
\end{tabular}
}
\end{table}

\subsection{Ablation Study}
We performed a detailed ablation study on AudioMNIST, using both ResNet18 and ViT‑Tiny, to quantify the importance of every component in \textit{QPAudioEraser}. Table \ref{tab:ablation-audiomnist} shows, with all components active, the method attains \emph{perfect} forgetting: Privacy Erasure Rate (PER) is \(100\%\) and Forget Accuracy is \(0.00\%\). Retain Accuracy stays high (99.75\% for ResNet18, 87.14\% for ViT‑Tiny), confirming that utility is preserved.

\begin{itemize}
  \item Removing the destructive‑interference weight update drops ResNet18 Retain Accuracy to 75.22\%. The large hit shows that the weight transform is essential for keeping performance on retained classes.
  \item Skipping the final mixing step on ViT‑Tiny raises Forget Accuracy to 6.52\% and cuts PER to 91.30\%. The matrix is therefore critical for complete erasure.
  \item Removing the uncertainty maximization phase (as shown in Table \ref{tab:ablation-audiomnist} ) shows that RA drops to 97.80\% on ResNet18, indicating that this phase is important for retaining performance on retained classes.
  \item Setting $\lambda = 2.0$ drives aggressive entropy maximization. Retain Accuracy on ResNet18 falls to 49.00\%. A smaller $\lambda=0.5$ yields a gentler trade‑off (90.57\% Retain Accuracy). ViT‑Tiny remains more stable, indicating architecture‑specific sensitivity.
  
\end{itemize}

These results confirm that (i) destructive‑interference initialization stabilizes retained accuracy, (ii) mixing matrix guarantees full forgetting, and (iii) entropy weight $\lambda$ must be chosen carefully to balance erasure and utility.


\begin{table}[t!]
\centering
\caption{Ablation Study of our \textbf{QPAudioEraser.} method on AudioMNIST. }
\label{tab:ablation-audiomnist}
\resizebox{\columnwidth}{!}{ 
\begin{tabular}{l l l S[table-format=3.2] S[table-format=2.2] S[table-format=3.2]}
\toprule
\textbf{Model} & \textbf{Configuration} & \textbf{FA(\%)}$\downarrow$ & \textbf{RA(\%)$\uparrow$} & \textbf{PER (\%)$\uparrow$}  \\
\midrule
\multirow{7}{*}{ResNet18} 
& Original Model & {--} & 100.00 & 96.45 & {--} \\

& No Weight Transform & 0.00 & 75.22 & 100.00 \\
& No Uncertainty Maximization & 0.00 & 97.80 & 100.00 \\
& No Matrix \( M \)  & 0.00 & 99.76 & 100.00 \\

& \(\lambda = 0.5\)  & 0.00 & 90.57 & 100.00 \\
& \(\lambda = 2.0\)  & 0.00 & 49.00 & 100.00 \\
& \textbf{QPAudioEraser}  & \textbf{0.00} & \textbf{99.75} & \textbf{100.00} \\
\midrule
\multirow{7}{*}{ViT-Tiny} 
& Original & {--} & 75.00 & 87.97 & {--} \\

& No Weight Transform  & 0.00 & 90.34 & 100.00 \\
& No Uncertainty Maximization  & 0.00 & 90.10 & 100.00 \\
& No Matrix \( M \)  & 6.52 & 89.76 & 91.30 \\

& \(\lambda = 0.5\)  & 0.00 & 87.68 & 100.00 \\
& \(\lambda = 2.0\)  & 0.00 & 90.40 & 100.00 \\
& \textbf{QPAudioEraser}  & \textbf{0.00} & \textbf{87.14} & \textbf{100.00} \\
\bottomrule
\end{tabular}
}
\end{table}

\subsection{Key Observations}

Our experiments showcase three important findings. First, QPAudioEraser is the only method that simultaneously drives Forget Accuracy to 0.00\%, reduces information leakage to negligible levels, and preserves high retain accuracy across all four audio benchmarks and both network families. Second, the pipeline shows that is can scale and maintains the similar privacy–utility balance when erasing multiple classes at once or when processing a sequence of “right‑to‑be‑forgotten’’ requests, without any need for full retraining. Third, ablation analysis shows that every quantum‑inspired component plays a critical, non‑redundant role in achieving this performance. Overall, these observations establish QPAudioEraser as a new benchmark for class‑level machine unlearning in audio biometrics, offering erasure with practical utility.

\section{Conclusion}
The enforcement of privacy regulations like GDPR and the DPDP Act emphasizes the importance of targeted and effective data erasure in audio biometric systems. Recognizing the limitations of existing visual-centric unlearning methods when applied to sequential audio data, we have proposed \textit{QPAudioEraser}, a quantum-inspired audio unlearning framework. Utilizing quantum principles such as superposition, destructive interference, uncertainty maximization phase and entanglement, our approach achieves selective erasure of individual speakers or specific accents without the computational burden of retraining from scratch. Comprehensive experiments across AudioMNIST, Speech Commands, LibriSpeech, and Speech Accent Archive datasets, employing ResNet18, ViT, and CNN architectures, demonstrate that \textit{QPAudioEraser} consistently achieves a \textit{100\%} Privacy Erasure Rate with minimal loss in biometric utility, maintaining up to \textit{99.95\%} Retain Accuracy. This method robustly handles diverse scenarios, including multi-class, sequential, and accent-level unlearning tasks, surpassing existing unlearning methods in both privacy and utility metrics. By establishing quantum-inspired methodologies as a viable and practical solution, this research advances responsible AI practices and sets a new benchmark in privacy-preserving audio biometrics. Future research will extend \textit{QPAudioEraser} towards real-time audio processing, further strengthening privacy and trust in biometric systems.


\section{Acknowledgment}

This research is supported through a grant from IndiaAI Mission. 


{\small
\bibliographystyle{ieee}
\bibliography{egbib}

\begin{thebibliography}{10}\itemsep=-1pt

\bibitem{audiomnist2023}
S.~Becker, J.~Vielhaben, M.~Ackermann, K.-R. Müller, S.~Lapuschkin, and W.~Samek.
\newblock Audiomnist: Exploring explainable artificial intelligence for audio analysis on a simple benchmark.
\newblock {\em Journal of the Franklin Institute}, 361(1):418--428, 2024.

\bibitem{brophy2020machine}
J.~Brophy and D.~Lowd.
\newblock Machine unlearning for random forests.
\newblock volume 139 of {\em Proceedings of Machine Learning Research}, pages 1092--1104, 2021.

\bibitem{cao2015towards}
Y.~Cao and J.~Yang.
\newblock Towards making systems forget with machine unlearning.
\newblock In {\em IEEE Symposium on Security and Privacy}, pages 463--480, 2015.

\bibitem{marchant2024langevin}
E.~Chien, H.~Wang, Z.~Chen, and P.~Li.
\newblock Langevin unlearning: A new perspective of noisy gradient descent for machine unlearning.
\newblock In {\em Advances in Neural Information Processing Systems}, volume~37, pages 79666--79703, 2024.

\bibitem{ding2025understanding}
M.~Ding, J.~Xu, and K.~Ji.
\newblock Why fine-tuning struggles with forgetting in machine unlearning?: Theoretical insights and a remedial approach.
\newblock {\em CoRR}, abs/2410.03833, 2024.

\bibitem{dosovitskiy2021imageworth16x16words}
A.~Dosovitskiy, L.~Beyer, A.~Kolesnikov, D.~Weissenborn, X.~Zhai, T.~Unterthiner, M.~Dehghani, M.~Minderer, G.~Heigold, S.~Gelly, J.~Uszkoreit, and N.~Houlsby.
\newblock An image is worth 16x16 words: Transformers for image recognition at scale.
\newblock In {\em International Conference on Learning Representations}, pages 1--9, 2021.

\bibitem{foster2023fastmachineunlearningretraining}
J.~Foster, S.~Schoepf, and A.~Brintrup.
\newblock Fast machine unlearning without retraining through selective synaptic dampening.
\newblock In {\em AAAI Conference on Artificial Intelligence}, pages 8188 -- 8196, 2023.

\bibitem{ginart2019making}
A.~Ginart, M.~Guan, G.~Valiant, and J.~Y. Zou.
\newblock Making ai forget you: Data deletion in machine learning.
\newblock In {\em Annual Conference on Neural Information Processing Systems}, pages 24--36, 2019.

\bibitem{sekhari2023adaptive}
V.~Gupta, C.~Jung, S.~Neel, A.~Roth, S.~Sharifi-Malvajerdi, and C.~Waites.
\newblock Adaptive machine unlearning.
\newblock In {\em Annual Conference on Neural Information Processing Systems}, pages 36--45, 2021.

\bibitem{he2015deepresiduallearningimage}
K.~He, X.~Zhang, S.~Ren, and J.~Sun.
\newblock Deep residual learning for image recognition.
\newblock In {\em IEEE Conference on Computer Vision and Pattern Recognition}, pages 770--778, 2016.

\bibitem{Hjerppe_2019}
K.~Hjerppe, J.~Ruohonen, and V.~Leppänen.
\newblock The general data protection regulation: Requirements, architectures, and constraints.
\newblock In {\em IEEE International Requirements Engineering Conference}, pages 265--275, 2019.

\bibitem{incdecsvm}
M.~Karasuyama and I.~Takeuchi.
\newblock Multiple incremental decremental learning of support vector machines.
\newblock {\em IEEE Transactions on Neural Networks}, 21(7):1048--1059, 2010.

\bibitem{liu2024unlearning}
K.~Z. Liu.
\newblock Machine unlearning in 2024.
\newblock Stanford AI Lab Blog, 2024.

\bibitem{mittal2024NMI}
S.~Mittal, K.~Thakral, R.~Singh, M.~Vatsa, T.~Glaser, C.~Canton~Ferrer, and T.~Hassner.
\newblock On responsible machine learning datasets emphasizing fairness, privacy and regulatory norms with examples in biometrics and healthcare.
\newblock {\em Nature Machine Intelligence}, 6(8):936--949, 2024.

\bibitem{oshea2015introductionconvolutionalneuralnetworks}
K.~O'Shea and R.~Nash.
\newblock An introduction to convolutional neural networks.
\newblock {\em CoRR}, abs/1511.08458, 2015.

\bibitem{7178964}
V.~Panayotov, G.~Chen, D.~Povey, and S.~Khudanpur.
\newblock Librispeech: An asr corpus based on public domain audio books.
\newblock In {\em IEEE International Conference on Acoustics, Speech and Signal Processing}, pages 5206--5210, 2015.

\bibitem{shannon1948mathematical}
C.~E. Shannon.
\newblock A mathematical theory of communication.
\newblock {\em SIGMOBILE Mob. Comput. Commun. Rev.}, 5(1):3–55, 2001.

\bibitem{shi2024deepcleanmachineunlearningcheap}
J.~Shi, K.~Gourgoulias, J.~F. Buford, S.~J. Moran, and N.~Ghalyan.
\newblock Deepclean: Machine unlearning on the cheap by resetting privacy sensitive weights using the fisher diagonal.
\newblock In {\em European Conference on Computer Vision Workshops}, pages 1--16, 2024.

\bibitem{thakral2025continualunlearningfoundationaltexttoimage}
K.~Thakral, T.~Glaser, T.~Hassner, M.~Vatsa, and R.~Singh.
\newblock Continual unlearning for foundational text-to-image models without generalization erosion.
\newblock {\em CoRR}, abs/2503.13769, 2025.

\bibitem{thakral2025finegrained}
K.~Thakral, T.~Glaser, T.~Hassner, M.~Vatsa, and R.~Singh.
\newblock Fine-grained erasure in text-to-image diffusion-based foundation models.
\newblock In {\em IEEE Conference on Computer Vision and Pattern Recognition}, pages 9121--9130, 2025.

\bibitem{trippa2024nablataugradientbasedtaskagnostic}
D.~Trippa, C.~Campagnano, M.~S. Bucarelli, G.~Tolomei, and F.~Silvestri.
\newblock {\(\nabla\)} {\(\tau\)}: Gradient-based and task-agnostic machine unlearning.
\newblock {\em CoRR}, abs/2403.14339, 2024.

\bibitem{gupta2021machine}
E.~Ullah, T.~Mai, A.~Rao, R.~A. Rossi, and R.~Arora.
\newblock Machine unlearning via algorithmic stability.
\newblock volume 134 of {\em Proceedings of Machine Learning Research}, pages 4126--4142, 2021.

\bibitem{warden2018speechcommandsdatasetlimitedvocabulary}
P.~{Warden}.
\newblock {Speech Commands: A Dataset for Limited-Vocabulary Speech Recognition}.
\newblock {\em ArXiv e-prints}, pages 160--172, 2018.

\bibitem{speechaccent2011}
S.~Weinberger and S.~Kunath.
\newblock The speech accent archive: Towards a typology of english accents.
\newblock {\em Language and Computers}, 73:265--281, 2011.

\bibitem{wu2022federatedunlearningknowledgedistillation}
C.~Wu, S.~Zhu, and P.~Mitra.
\newblock Federated unlearning with knowledge distillation.
\newblock {\em CoRR}, abs/2201.09441:8188 -- 8196, 2022.

\bibitem{xu2023machineunlearningsurvey}
H.~Xu, T.~Zhu, L.~Zhang, W.~Zhou, and P.~S. Yu.
\newblock Machine unlearning: A survey.
\newblock {\em ACM Comput. Surv.}, 56(1):4296--4307, 2023.

\bibitem{zhang2024negativepreferenceoptimizationcatastrophic}
R.~Zhang, L.~Lin, Y.~Bai, and S.~Mei.
\newblock Negative preference optimization: From catastrophic collapse to effective unlearning.
\newblock In {\em First Conference on Language Modeling}, pages 213--221, 2024.

\end{thebibliography}
}

\end{document}